\documentclass[aps,preprint,prb,showpacs,floatfix,superscriptaddress]{revtex4}

\usepackage{epsfig,amsmath,amssymb} 
\usepackage{graphics}
\usepackage{multirow}
\begin{document}

\title
{
Magnetic order in a spin-half interpolating square-triangle Heisenberg antiferromagnet
}
\author
{
R.~F.~Bishop 
}
\affiliation
{
School of Physics and Astronomy, Schuster Building, The University of Manchester, Manchester, M13 9PL, UK}

\author
{
P.~H.~Y.~Li
}
\affiliation
{
School of Physics and Astronomy, Schuster Building, The University of Manchester, Manchester, M13 9PL, UK}

\author 
{
D.~J.~J.~Farnell
} 
\affiliation {Academic Department of Radiation Oncology, The University of Manchester, The Christie NHS Foundation Trust, Wilmslow Road, Manchester M20 4BX, UK}

\author
{
C.~E.~Campbell
}
\affiliation
{
School of Physics and Astronomy, University of Minnesota, 116 Church Street SE, Minneapolis, Minnesota 55455, USA}

\begin{abstract}
  Using the coupled cluster method (CCM) we study the zero-temperature
  phase diagram of a spin-half Heisenberg antiferromagnet (HAF), the
  so-called $J_{1}$--$J_{2}'$ model, defined on an anisotropic
  two-dimensional lattice.  With respect to an underlying
  square-lattice geometry the model contains antiferromagnetic ($J_{1}
  > 0$) bonds between nearest neighbors and competing ($J_{2}'>0$)
  bonds between next-nearest neighbors across only one of the
  diagonals of each square plaquette, the same diagonal in every
  square.  Considered on an equivalent triangular-lattice geometry the
  model may be regarded as having two sorts of nearest-neighbor bonds,
  with $J_{2}' \equiv \kappa J_{1}$ bonds along parallel chains and
  $J_{1}$ bonds providing an interchain coupling.  Each triangular
  plaquette thus contains two $J_{1}$ bonds and one $J_{2}'$ bond.
  Hence, the model interpolates between a spin-half HAF on the square
  lattice at one extreme ($\kappa = 0$) and a set of decoupled
  spin-half chains at the other ($\kappa \rightarrow \infty$), with
  the spin-half HAF on the triangular lattice in between at $\kappa =
  1$.  We use a N\'{e}el state, a helical state, and a collinear
  stripe-ordered state as separate starting model states for the CCM
  calculations that we carry out to high orders of approximation (up
  to LSUB$8$).  The interplay between quantum fluctuations, magnetic
  frustration, and varying dimensionality leads to an interesting
  quantum phase diagram.  We find strong evidence that quantum
  fluctuations favor a weakly first-order or possibly second-order
  transition from N\'{e}el order to a helical state at a first
  critical point at $\kappa_{c_{1}} = 0.80 \pm 0.01$, by contrast with
  the corresponding second-order transition between the equivalent
  classical states at $\kappa_{{\rm cl}} = 0.5$.  We also find strong
  evidence for a second critical point at $\kappa_{c_{2}} = 1.8 \pm
  0.4$ where a first-order transition occurs, this time from the
  helical phase to a collinear stripe-ordered phase.  This latter
  result provides quantitative verification of a recent qualitative
  prediction of Starykh and Balents [Phys.\ Rev.\ Lett. {\bf 98},
  077205 (2007)] based on a renormalization group analysis of the
  $J_{1}$--$J_{2}'$ model that did not, however, evaluate the
  corresponding critical point.
\end{abstract}

\pacs{75.10.Jm, 75.30.Gw, 75.40.-s, 75.50.Ee}

\maketitle

\section{Introduction}
\label{Introd}
Two-dimensional (2D), spin-1/2, Heisenberg antiferromagnets (HAFs)
have been much studied in recent years.  The interplay between (either
dynamic or geometric) frustration and quantum fluctuations in
determining the ground-state (gs) phase diagram of such models has
been of particular interest.  While such models are well understood in
the absence of frustration,~\cite{Sa:1995} this is not the case for
frustrated systems, for which the zero-temperature ($T = 0$) phase
transitions between magnetically ordered quasiclassical phases and
novel (magnetically disordered) quantum paramagnetic
phases~\cite{Ri:2004,Mi:2005} have become the subject of great recent
interest.  A particularly well studied such model is the frustrated
$J_{1}$--$J_{2}$ model on the square lattice with nearest-neighbor
(NN) bonds ($J_{1}$) and next-nearest-neighbor (NNN) bonds ($J_{2}$),
for which it is now well accepted that there exist two phases
exhibiting magnetic long-range order (LRO) at small and
at large values of $\alpha \equiv J_{2}/J_{1}$ respectively, separated
by an intermediate quantum paramagnetic phase without magnetic LRO in
the parameter regime $\alpha_{c_{1}} < \alpha < \alpha_{c_{2}}$, where
$\alpha_{c_{1}} \approx 0.4$ and $\alpha_{c_{2}} \approx 0.6$.  For
$\alpha < \alpha_{c_{1}}$ the gs phase exhibits N\'{e}el magnetic LRO,
whereas for $\alpha > \alpha_{c_{2}}$ it exhibits collinear stripe
LRO.  We have recently studied this 2D spin-1/2 model exhaustively by
extending it to include anisotropic interactions in either real
(crystal lattice) space~\cite{Bi:2008_JPCM} or in spin
space.~\cite{Bi:2008_PRB}  We showed in particular how the coupled
cluster method (CCM) provided for this highly frustrated model what is
perhaps now the most accurate microscopic description.  The interested
reader is referred to Refs.~[\onlinecite{Bi:2008_JPCM,Bi:2008_PRB}] and
references cited therein for further details of the model and the
method.

\section{The model}
\label{model_section}
\begin{figure}[t]
%\begin{center}
%\epsfig{file=fig1.eps,width=6cm}
\epsfig{file=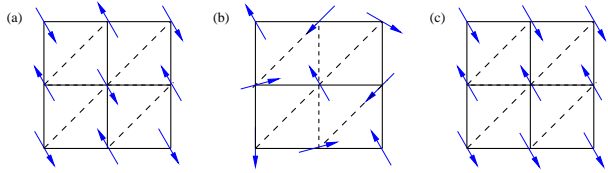,width=8cm}
\caption{(Color online) $J_{1}$--$J_{2}'$ model; --- $J_{1}$; - - -
  $J_{2}'$; (a) N\'{e}el state, (b) spiral state, (c) stripe state.}
\label{model}
%\end{center}
\end{figure}

In the light of the above successes we now apply the CCM to the
seemingly similar 2D spin-1/2 $J_{1}$--$J_{2}'$ model that has been
studied recently by other means.~\cite{Ga:1993,Tr:1999,Me:1999,We:1999,St:2007,Pa:2008}  Its
Hamiltonian is written as
\begin{equation}
H = J_{1}\sum_{\langle i,j \rangle}{\bf s}_{i}\cdot{\bf s}_{j} + J_{2}'\sum_{[i,k]}{\bf s}_{i}\cdot {\bf s}_{k}  \label{H}
\end{equation}
where the operators ${\bf s}_{i} \equiv (s^{x}_{i}, s^{y}_{i},
s^{z}_{i})$ are the spin operators on lattice site $i$ with ${\bf
  s}^{2}_{i} = s(s+1)$ and $s=1/2$.  On the square lattice the sum
over $\langle i,j \rangle$ runs over all distinct NN bonds, but the
sum over $[i,k]$ runs only over one half of the distinct NNN bonds
with equivalent bonds chosen in each square plaquette, as shown
explicitly in Fig.~\ref{model}.
(By contrast, the $J_{1}$--$J_{2}$ model discussed above includes {\it
  all} of the diagonal NNN bonds.)  We shall be interested here only
in the case of competing (or frustrating) antiferromagnetic bonds
$J_{1} > 0$ and $J_{2}' > 0$, and henceforth for all of the results
shown we set $J_{1} \equiv 1$.  Clearly, the model may be described
equivalently as a Heisenberg model on an anisotropic triangular
lattice in which each triangular plaquette contains two NN $J_{1}$
bonds and one NN $J_{2}'$ bond.  The model thus interpolates
continuously between HAFs on a square lattice ($J_{2}' = 0$) and on a
triangular lattice ($J_{2}'=J_{1}$).  Similarly, when $J_{1}=0$ (or
$J_{2}' \rightarrow \infty$ in our normalization with $J_{1} \equiv
1$) the model reduces to uncoupled 1D chains (along the chosen
diagonals on the square lattice).  The case $J_{2}' \gg 1$ thus
corresponds to weakly coupled 1D chains, and hence also interpolates
between 1D and 2D.  We note in this context that the CCM has also been
very successfully applied to other spin-1/2 HAF models that
continuously interpolate between (a) the triangular and kagom\'{e}
lattices;~\cite{Fa:2001} (b) the square and honeycomb
lattices;~\cite{Kr:2000} (c) 1D and 2D cases;~\cite{Bi:2008_JPCM} and
(d) 2D and 3D cases.~\cite{Schm:2006}  As well as the obvious
theoretical richness of the model, there is also experimental interest
since it also well describes such quasi-2D materials as BEDT-TTF
crystals~\cite{Ki:1996} with $J_{2}'/J_{1} \approx 0.34$--$1$, and
Cs$_{2}$CuCl$_{4}$~\cite{Co:1997} with $J_{2}'/J_{1} \approx 6$.

The $J_{1}$--$J_{2}'$ model has only two classical gs phases
(corresponding to the case where the spin quantum number $s
\rightarrow \infty$).  For $J_{2}' < \frac{1}{2}J_{1}$ the gs phase is N\'{e}el
ordered, as shown in Fig.~\ref{model}(a), whereas for $J_{2}' >
\frac{1}{2}J_{1}$ it has spiral order, as shown in Fig.~\ref{model}(b),
wherein the spin direction at lattice site ($i,j$) points at an angle
$\alpha_{ij}=\alpha_{0}+(i+j)\alpha_{{\rm cl}}$, with $\alpha_{{\rm
    cl}}={\rm cos}^{-1}(-\frac{J_{1}}{2J_{2}'}) \equiv \pi -
\phi_{{\rm cl}}$.  The pitch angle $\phi_{{\rm cl}}={\rm
  cos}^{-1}(\frac{J_{1}}{2J_{2}'})$ thus measures the deviaton from
N\'{e}el order, and it varies from zero for $2J_{2}'/J_{1} \leq 1$ to
$\frac{1}{2} \pi$ as $J_{2}'/J_{1} \rightarrow \infty$, as shown in
Fig.~\ref{angleVSj2}.  When $J_{2}'=J_{1}$ we regain the classical
3-sublattice ordering on the triangular lattice with $\alpha_{{\rm
    cl}} = \frac{2}{3}\pi$.  The classical phase transition at
$J_{2}'=\frac{1}{2}J_{1}$ is of continuous (second-order) type, with
the gs energy and its derivative both continuous.

In the limit of large $J_{2}'/J_{1}$ the above classical limit
represents a set of decoupled 1D HAF chains (along the diagonals of
the square lattice) with a relative spin orientation between
neighboring chains that approaches 90$^{\circ}$.  In fact, of course,
there is complete degeneracy at the classical level in this limit
between all states for which the relative ordering directions of spins
on different HAF chains are arbitrary.  Clearly the exact spin-1/2
limit should also be a set of decoupled HAF chains as given by the
exact Bethe ansatz solution.~\cite{Be:1931} However, one might expect
that this degeneracy could be lifted by quantum fluctuations by the
well-known phenomenon of {\it order by disorder}.~\cite{Vi:1977} Just
such a phase is known to exist in the $J_{1}$--$J_{2}$
model~\cite{Bi:2008_JPCM,Bi:2008_PRB} for values of $J_{2}/J_{1}
\gtrsim 0.6$, where it is the so-called collinear stripe phase in
which, on the square lattice, spins along (say) the rows in
Fig.~\ref{model} order ferromagnetically while spins along the columns
and diagonals order antiferromagnetically, as shown in Fig.\
\ref{model}(c).  We note, however, that a corresponding order by disorder phenomenon, if it exists for the present $J_{1}$--$J_{2}'$ model, would be more subtle than for its textbook $J_{1}$--$J_{2}$ model counterpart, as we explain more fully in Sec.\ V.  

In a recent paper Starykh and Balents~\cite{St:2007}
have given a renormalization group (RG) analysis of the spin-1/2
$J_{1}$--$J_{2}'$ model considered here to predict that precisely such a collinear
stripe phase also exists in this case for values of $J_{2}'/J_{1}$
above some critical value (which they do not calculate).  One of the
aims of the present paper is to give a fully microscopic analysis of
this model in order to map out its $T=0$ phase diagram, including the
positions and orders of any quantum phase transitions that emerge.

\section{The coupled cluster method}
The CCM (see, e.g., Refs.~[\onlinecite{Bi:1991,Bi:1998,Fa:2004}] and
references cited therein) that we employ here is one of the most
powerful and most versatile modern techniques in quantum many-body
theory.  It has been applied very successfully to various quantum
magnets (see
Refs.~[\onlinecite{Bi:2008_JPCM,Bi:2008_PRB,Fa:2001,Kr:2000,Schm:2006,Fa:2004,Ze:1998,Da:2005}]
and references cited therein).  The method is particularly appropriate
for studying frustrated systems, for which the main alternative
methods are often only of limited usefulness.  For example, quantum
Monte Carlo techniques are particularly plagued by the sign problem
for such systems, and the exact diagonalization method is restricted
in practice, particularly for $s>1/2$, to such small lattices that it
is often insensitive to the details of any subtle phase order present.

The method of applying the CCM to quantum magnets has been described
many times elsewhere (see, e.g.,
Refs.~[\onlinecite{Fa:2001,Kr:2000,Schm:2006,Bi:1991,Bi:1998,Fa:2004}]
and references cited therein).  It relies on building multispin
correlations on top of a chosen gs model state $|\Phi\rangle$ in a
systematic hierarchy of LSUB$n$ approximations for the correlation
operators $S$ and $\tilde{S}$ that exactly parametrize the exact gs
ket and bra wave functions of the system respectively as $|\Psi
\rangle=e^{S}|\Phi\rangle$ and $\langle \tilde{\Psi}|=\langle
\Phi|\tilde{S}e^{-S}$.  In the present case we use three different
choices for the model state $|\Phi\rangle$, namely either of the
classical N\'{e}el and spiral states, as well as the collinear
stripe state.  Note that for the helical phase we perform calculations
for arbitrary pitch angle $\alpha \equiv \pi - \phi$, and then
minimize the corresponding LSUB$n$ approximation for the energy with
respect to $\phi$, $E_{{\rm LSUB}n}(\phi) \rightarrow {\rm min}
\Leftrightarrow \phi = \phi_{{\rm LSUB}n}$.  Generally (for $n > 2$)
the minimization must be carried out computationally in an iterative
procedure, and for the highest values of $n$ that we use here the use
of supercomputing resources was essential.  Results for $\phi_{{\rm
    LSUB}n}$ will be given later (Fig.~\ref{angleVSj2}).  We choose
local spin coordinates on each site in each case so that all spins in
$|\Phi\rangle$, whatever the choice, point in the negative
$z$-direction (i.e., downwards).

Then, in the LSUB$n$ approximation all possible multi-spin-flip
correlations over different locales on the lattice defined by $n$ or
fewer contiguous lattice sites are retained.  Clearly, in the present
case we have a choice whether to consider the model to be defined on
the square lattice (shown in Fig.\ \ref{model}) or to consider it on
the (topologically equivalent) triangular lattice, as discussed in
Sec.\ \ref{model_section}.  Although these two viewpoints are
completely equivalent for a description of the model, they differ for
the purposes of defining the LSUB$n$ approximations.  Thus, for
example, each pair of sites joined by a $J_{2}'$ bond are NNN pairs on
the square lattice but are NN pairs on the triangular lattice.  Hence,
such a (NN on the triangular lattice) double spin-flip configuration
is contained in LSUB$n$ approximations on the square lattice only for
$n \geq 3$, whereas it is contained at the LSUB$n$ level on the
triangular lattice for $n \geq 2$.  Whereas both LSUB$n$ hierachies
agree in the $n \rightarrow \infty$ limit they will differ for finite
values of $n$.  In general there are clearly more multi-spin-flip
configurations retained at a given LSUB$n$ level on the triangular
lattice than on the square lattice, and in the present paper we
consider only the triangular case.

The numbers of such distinct fundamental configurations on the
triangular lattice (viz., those that are distinct under the space and
point group symmetries of both the Hamiltonian and the model state
$|\Phi\rangle$) that are retained for the collinear stripe and spiral
states of the current model in various LSUB$n$ approximations are
shown in Table~\ref{table_FundConfig}.
\begin{table}[t]
\caption{Number of fundamental LSUB$n$ configurations ({$\sharp$ f.c.}) for the stripe and 
spiral states of the spin-$1/2$ $J_{1}$--$J_{2}'$ model, using the triangular lattice geometry.}
\label{table_FundConfig}
\begin{tabular}{ccc} \hline\hline
\multirow{2}{*}{Method} & \multicolumn{2}{c}{$\sharp$ f.c.} \\ \cline{2-3}
& stripe & spiral \\ \hline
LSUB$2$ & 2 & 3 \\ 
LSUB$3$ & 4 & 14 \\ 
LSUB$4$ & 27 & 67 \\ 
LSUB$5$ & 95  & 370 \\ 
LSUB$6$ & 519 & 2133\\ 
LSUB$7$ & 2617 & 12878 \\ 
LSUB$8$ & 15337 & 79408 \\ \hline\hline
\end{tabular} 
\end{table}
The coupled sets of equations for these corresponding numbers of
coefficients in the operators $S$ and $\tilde{S}$ are derived using
computer algebra~\cite{ccm} and then solved~\cite{ccm} using parallel
computing.  We note that such CCM calculations using up to about
$10^{5}$ configurations or so have been previously carried out many
times using the CCCM code\cite{ccm} and heavy parallelization.  A
significant extra computational burden arises here for the helical
state due to the need to optimize the quantum pitch angle at each
LSUB$n$ level of approximation as described above.  Furthermore, for
many model states the quantum number $s^{z}_{T} \equiv \sum^{N}_{i=1}
s^{z}_{i}$ may be used to restrict the numbers of fundamental
multi-spin-flip configurations to those clusters that preserve
$s^{z}_{T}=0$. However, for the spiral model state that symmetry is
absent, which largely explains the significantly greater number of
fundamental configurations for the spiral state than for the stripe
state at a given LSUB$n$ order.  Hence, the maximum LSUB$n$ level that
we can reach here, even with massive parallelization and the use of
supercomputing resources, is LSUB$8$.  For example, to obtain a single
data point (i.e., for a given value of $J_{2}'$, with $J_{1}=1$) for
the spiral phase at the LSUB$8$ level typically required about 0.3 h
computing time using 600 processors simultaneously. 

At each level of approximation we may then calculate a
corresponding estimate of the gs expectation value of any physical
observable such as the energy $E$ and the magnetic order parameter, $M
\equiv -\langle \tilde{\Psi}|s^{z}_{i}|\Psi \rangle$, defined in the
local, rotated spin axes, and which thus represents the on-site
magnetization.  Note that $M$ is just the usual sublattice
magnetization for the case of the N\'{e}el state as the CCM model
state, for example.  More generally it is just the on-site
magnetization.

It is important to note that we never need to perform any finite-size
scaling, since all CCM approximations are automatically performed from
the outset in the infinite-lattice limit, $N \rightarrow \infty$,
where $N$ is the number of lattice sites.  However, we do need as a
last step to extrapolate to the $n \rightarrow \infty$ limit in the
LSUB$n$ truncation index $n$.  We use here the
well-tested~\cite{Fa:2001,Kr:2000} empirical scaling laws
\begin{equation}
E/N=a_{0}+a_{1}n^{-2}+a_{2}n^{-4}\;,  \label{Extrapo_E}
\end{equation} 
\begin{equation}
M=b_{0}+b_{1}n^{-1}+b_{2}n^{-2}\;, \label{Extrapo_M}
\end{equation} 
that have given good results previously, for example, for the
interpolating triangle-kagom\'{e} HAF~\cite{Fa:2001} and the
interpolating square-honeycomb HAF.\cite{Kr:2000}  We comment further on the accuracy of the extrapolations in Sec.\ \ref{discussion} where we present a discussion of our results.

\section{Results}
We report here on CCM calculations for the present spin-1/2
$J_{1}$--$J_{2}'$ model Hamiltonian of Eq.\ (\ref{H}) for given
parameters ($J_{1}=1$, $J_{2}'$), based respectively on the N\'{e}el,
spiral and stripe states as CCM model states.  Our computational power
is such that we can perform LSUB$n$ calculations for each model state
with $n \leq 8$.  We note that, as has been well documented in the
past,~\cite{Fa:2008} the LSUB$n$ data for both the gs energy per spin
$E/N$ and the on-site magnetization $M$ converge differently for the
even-$n$ sequence and the odd-$n$ sequence, similar to what is
frequently observed in perturbation theory.~\cite{Mo:1953} Since, as a
general rule, it is desirable to have at least ($n+1$) data points to
fit to any fitting formula that contains $n$ unknown parameters, we
prefer to have at least 4 results to fit to Eqs.\ (\ref{Extrapo_E})
and (\ref{Extrapo_M}).  Hence, for most of our extrapolated results
below we use the even LSUB$n$ sequence with $n=\{2,4,6,8\}$.

\begin{figure}[t]
%\begin{center}
%\epsfig{file=fig2.eps,width=5.5cm,angle=270}
\epsfig{file=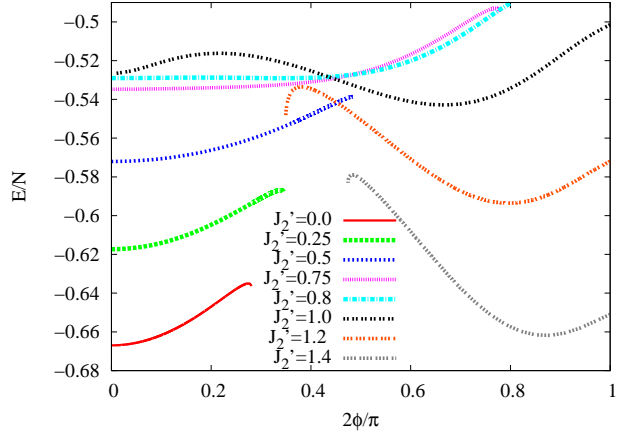,width=6cm,angle=270}
\caption{(Color online) Ground-state energy per spin of the spin-1/2
  $J_{1}$--$J_{2}'$ Hamiltonian of Eq.\ (\ref{H}) with $J_{1}=1$,
  using the LSUB6 approximation of the CCM with the spiral model
  state, versus the spiral angle $\phi$, for some illustrative values
  of $J_{2}'$ in the range $0 \leq J_{2}' \leq 1.4$.  For $J_{2}'
  \lesssim 0.788$ the minimum is at $\phi=0$ (N\'{e}el order), whereas
  for $J_{2}' \gtrsim 0.788$ the minimum occurs at $\phi=\phi_{{\rm
      LSUB}6} \neq 0$, indicating a phase transition at $J_{2}'
  \approx 0.788$ in this approximation.}
\label{EvsAngle}
%\end{center}
\end{figure}

We report first on results obtained using the spiral model state.
While classically we have a second-order phase transition from
N\'{e}el order (for $\kappa < \kappa_{{\rm cl}}$) to helical order
(for $\kappa > \kappa_{{\rm cl}}$), where $\kappa \equiv
J_{2}'/J_{1}$, at a value $\kappa_{{\rm cl}} = 0.5$, using the CCM we
find strong indications of a shift of this critical point to a value
$\kappa_{c_{1}} \approx 0.80$ in the spin-1/2 quantum case.  Thus, for
example, curves such as those shown in Fig.\ \ref{EvsAngle}
show that the N\'{e}el model state ($\phi=0$) gives the minimum gs
energy for all values of $\kappa < \kappa_{c_{1}}$ where
$\kappa_{c_{1}}$ is also dependent on the level of LSUB$n$
approximation, as we also see below in Fig.\ \ref{angleVSj2}.
\begin{figure}[t]
%\begin{center}
%\epsfig{file=fig3.eps,width=5.5cm,angle=270}
\epsfig{file=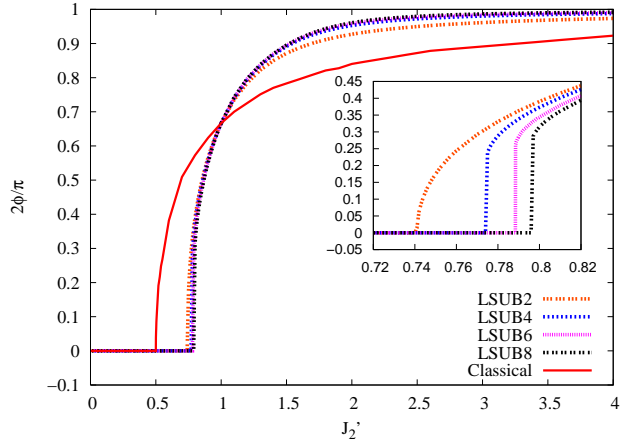,width=6cm,angle=270}
\caption{(Color online) The angle $\phi_{{\rm LSUB}n}$ that minimizes
  the energy $E_{{\rm LSUB}n}(\phi)$ of the spin-1/2 $J_{1}$--$J_{2}'$
  Hamiltonian Eq.\ (\ref{H}) with $J_{1}=1$, in the LSUB$n$
  approximations with $n=\{2,4,6,8\}$, using the spiral model state,
  versus $J_{2}'$.  The corresponding classical result $\phi_{{\rm
      cl}}$ is shown for comparison.  We find in the LSUB$n$ quantum
  case with $n > 2$ a seemingly first-order phase transition (e.g., for
  LSUB8 at $J_{2}' \approx 0.796$ where $\phi_{{\rm LSUB}8}$ jumps
  abruptly from zero to about $0.14 \pi$), although also see
  the text for a broader discussion of the nature of the quantum phase
  transition.  By contrast, in the classical case there is a
  second-order phase transition at $J_{2}'=0.5$.}
\label{angleVSj2}
%\end{center}
\end{figure}
By contrast, for $\kappa > \kappa_{c_{1}}$ the minimum in the energy
is found to occur at a value $\phi \neq 0$.  If we consider the pitch
angle $\phi$ itself as an order parameter (i.e., $\phi=0$ for N\'{e}el
order and $\phi \neq 0$ for spiral order) a typical scenario for a
first-order phase transition would be the appearance of a two-minimum
structure for the ground-state energy as a function of $\phi$, exactly
as shown in Fig.\ \ref{EvsAngle} for the LSUB6 approximation.  Very
similar curves occur for other LSUB$n$ approximations.  If we therefore admit
such a scenario, in the typical case one would expect various special
points in the transition region, namely the phase transition point
$\kappa_{c_{1}}$ itself where the two minima have equal depth, plus
one or two instability points $\kappa_{i_{1}}$ and $\kappa_{i_{2}}$
where one or other of the minima (at $\phi = 0$ and $\phi \neq 0$
respectively) disappears.  In the present case, it is interesting to
note that the two points $\kappa_{c_{1}}$ and $\kappa_{i_{2}}$ either
coincide exactly or are indistinguishable within the accuracy of our
calculations, thereby indicating that the transition at
$\kappa_{c_{1}}$ is rather subtle, and perhaps even of second-order
rather than first-order type.  A close inspection of curves such as those
shown in Fig.\ \ref{EvsAngle} for the LSUB6 case shows that what
happens at this level of approximation is that for $\kappa \lesssim
0.788$ the only minimum in the ground-state energy is at $\phi=0$
(N\'{e}el order).  As this value is approached asymptotically from
below the LSUB6 energy curves become extremely flat near $\phi \neq
0$, indicating the disappearance at $\phi=0$ of the second derivative
$d^{2}E/d\phi^{2}$ (and possibly also of one or more of the higher
derivatives $d^{n}E/d\phi^{n}$ with $n \geq 3$), as well as of the
first derivative $dE/d\phi$.  Then, for all values $\kappa \gtrsim
0.788$ the LSUB6 curves develop a secondary minimum at a value $\phi
\neq 0$ which is also the global minimum.

The state for $\phi \neq 0$ is believed to be the quantum analog of
the classical spiral phase.  The fact that N\'{e}el order survives
beyond the classically stable region is an example of the promotion of
collinear order by quantum fluctuations, a phenomenon that has been
observed in many other systems (see, e.g.,
Refs.~[\onlinecite{Kr:2000,Ko:1996}]).  Thus, this collinear ordered
state survives for the quantum case into a region where classically it
is already unstable.  Indeed, one can view this behavior more broadly
as another example of the more general phenomenon of order by
disorder~\cite{Vi:1977} that we have briefly alluded to above, in
which quantum fluctuations act to select and stabilize an appropriate
type of order (that is typically collinear) in the face of classical
degeneracy or near-degeneracy.

It is also particularly interesting to note that the crossover from
one minimum ($\phi=0$, N\'{e}el) solution to the other ($\phi \neq 0$,
spiral) appears (except for the LSUB$2$ case) to be quite abrupt (for
all other LSUB$n$ cases with even $n > 2$) at this point (and see
Figs.\ \ref{EvsAngle} and \ref{angleVSj2}).  Thus, for example, for
the LSUB6 case the spiral pitch angle $\phi$ appears to jump
discontinuously from a zero value on the N\'{e}el side ($\kappa
\lesssim 0.788$) to a value of about $0.13\pi$ as the transition point
into the spiral phase is crossed.  This behavior is a clear first
indication of a phase transition.  Based on this evidence alone it
would also appear that this transition is first-order, by contrast
with the second-order nature of its classical counterpart.  Such a
situation where the quantum fluctuations change the nature of a phase
transition qualitatively from a classical second-order type to a
quantum first-order type has also been seen previously in the
comparable spin-1/2 HAF model that interpolates continuously between
the square and honeycomb lattices.~\cite{Kr:2000} However, due to the
extreme insensitivity of the energy to the pitch angle near the phase
transition, as discussed above, we cannot rule out a continuous but
very steep rise in pitch angle as the transition from the N\'{e}el
phase into the spiral phase is transversed.  All of the available
evidence to date indicates that the transition at $\kappa_{c_{1}}$ is
subtle and may actually be second-order.  Further evidence for the
position $\kappa_{c_{1}}$ and nature of the N\'{e}el--spiral quantum
phase transition also comes from the behavior of the on-site
magnetization that we discuss below.

Before doing so, however, we wish to make some further observations on
Figs.\ \ref{EvsAngle} and \ref{angleVSj2}.  We note first from Fig.\
\ref{angleVSj2} that in the case $\kappa=1$ ($J_{1}=1, J_{2}'=1$),
corresponding to the spin-1/2 HAF on the triangular lattice, all of the
CCM LSUB$n$ approximations give precisely the classical value
$\phi=\frac{\pi}{3}$ for the spiral angle, which corresponds to the
correct 120$^{\circ}$ three-sublattice ordering.  The fact that all
LSUB$n$ approximations give exactly this value is a consequence of us
defining the LSUB$n$ configurations on the triangular lattice (rather
than the square lattice), and is a reflection of the exact triangular
symmetry that is thereby preserved by our approximations.  It is also
interesting to note that for values of $\kappa > 1$ the quantum spiral
angle $\phi$ approaches the asymptotic ($\kappa \rightarrow \infty$)
value of $\frac{\pi}{2}$ much faster than does the classical angle.  This
is a first indication again, in this limit, of quantum fluctuations
favoring collinear order (along the weakly coupled chains in this
limit).

We note from Fig.\ \ref{EvsAngle} that for certain values of $J_{2}'$
(or, equivalently, $\kappa$) CCM solutions at a given LSUB$n$ level of
approximation (viz., LSUB$6$ in Fig.\ \ref{EvsAngle}) exist only for
certain ranges of spiral angle $\phi$.  For example, for the pure
square-lattice HAF ($\kappa=0$) the CCM LSUB$6$ solution based on a
spiral model state only exists for $0 \leq \phi \lesssim 0.14 \pi$.
In this case, where the N\'{e}el solution is the stable ground state,
if we attempt to move too far away from N\'{e}el collinearity the CCM
equations themselves become ``unstable'' and simply do not have a real
solution.  Similarly, we see from Fig.\ \ref{EvsAngle} that for
$\kappa=1.4$ the CCM LSUB$6$ solution exists only for $0.24 \pi
\lesssim \phi \leq 0.5 \pi$.  In this case the stable ground state is
a spiral phase, and now if we attempt to move too close to N\'{e}el
collinearity the real solution terminates.

Such terminations of CCM solutions are very common and are very well
documented.\cite{Fa:2004} In all such cases a termination point always
arises due to the solution of the CCM equations becoming complex at
this point, beyond which there exist two branches of entirely
unphysical complex conjugate solutions.\cite{Fa:2004} In the region
where the solution reflecting the true physical solution is real there
actually also exists another (unstable) real solution.  However, only
the (shown) upper branch of these two solutions reflects the true
(stable) physicsl ground state, whereas the lower branch does not.
The physical branch is usually easily identified in practice as the
one which becomes exact in some known (e.g., perturbative) limit.
This physical branch then meets the corresponding unphysical branch at
some termination point (with infinite slope on Fig.\ \ref{EvsAngle})
beyond which no real solutions exist.  The LSUB$n$ termination points
are themselves also reflections of the quantum phase transitions in
the real system, and may be used to estimate the position of the phase
boundary,\cite{Fa:2004} although we do not do so for this first
critical point since we have more accurate criteria discussed below.

\begin{figure}[t]
\epsfig{file=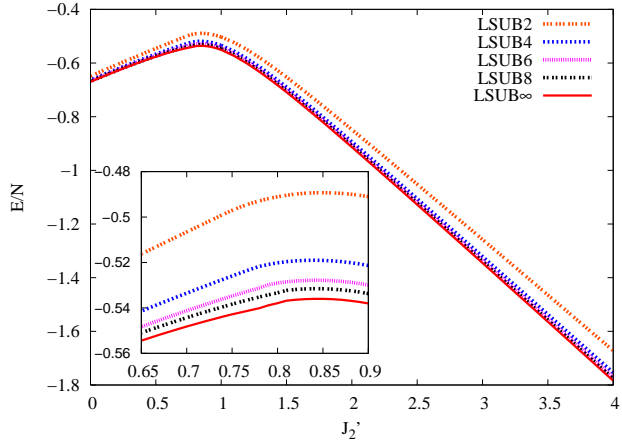,width=6cm,angle=270}
\caption{(Color online) Ground-state energy per spin versus $J_{2}'$
   for the N\'{e}el and spiral phases of the spin-1/2 $J_{1}$--$J_{2}'$ Hamiltonian of Eq.\ (\ref{H})
  with $J_{1}=1$.  The CCM results
  using the spiral model state are shown for various LSUB$n$
  approximations ($n=\{2,4,6,8\}$) with the spiral angle
  $\phi=\phi_{{\rm LSUB}n}$ that minimizes $E_{{\rm LSUB}n}(\phi)$.
  We also show the $n \rightarrow \infty$ extrapolated result from
  using Eq.\ (\ref{Extrapo_E}).}
\label{E}
\end{figure}

\begin{figure}[b]
%\begin{center}
%\epsfig{file=fig5.eps,width=5.5cm,angle=270}
\epsfig{file=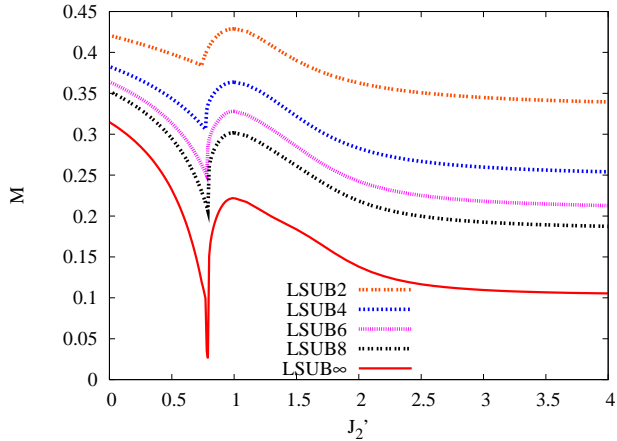,width=6cm,angle=270}
\caption{(Color online) Ground-state magnetic order parameter (i.e.,
  the on-site magnetization) versus $J_{2}'$ for
  the N\'{e}el and spiral phases of the spin-1/2
  $J_{1}$--$J_{2}'$ Hamiltonian of Eq.\ (\ref{H}) with $J_{1}=1$.  The CCM results using the spiral
  model state are shown for various LSUB$n$ approximations
  ($n=\{2,4,6,8\}$) with the spiral angle $\phi=\phi_{{\rm LSUB}n}$
  that minimizes $E_{{\rm LSUB}n}(\phi)$.  We also show the $n
  \rightarrow \infty$ extrapolated result from using Eq.\
  (\ref{Extrapo_M}).}
\label{M}
%\end{center}
\end{figure}

Thus, in Figs.\ \ref{E} and \ref{M} we show the CCM results for the gs
energy and gs on-site magnetization, respectively, where the helical
state has been used as the model state and the angle $\phi$ chosen as
described above.
%%%%%%%%%%%%%%%%%%%
For both quantities we show the raw LSUB$n$ data for $n=\{2,4,6,8\}$
and the extrapolated (LSUB$\infty$) results obtained from them by
using Eqs.\ (\ref{Extrapo_E}) and (\ref{Extrapo_M}) respectively.
Firstly, the gs energy (in Fig.\ \ref{E}) shows signs of a (weak)
discontinuity in slope at the critical values $\kappa_{c_{1}}$
discussed above.  These values for $\kappa_{c_{1}}$ themselves depend
weakly on the approximation level.

Secondly, the gs magnetic order parameter in Fig.\ \ref{M} shows much
stronger and much clearer evidence of a phase transition at the
corresponding $\kappa_{c_{1}}$ values previously observed in Fig.\
\ref{angleVSj2}.  The extrapolated value of $M$ shows clearly its
steep drop towards a value very close to zero at $\kappa_{c_{1}}=0.80
\pm 0.01$, which is hence our best estimate of the phase transition
point.  From the N\'{e}el side ($\kappa < \kappa_{c_{1}}$) the
magnetization seems to approach continuously a value $M=0.025 \pm
0.025$, whereas from the spiral side ($\kappa > \kappa_{c_{1}}$) there
appears to be a discontinuous jump in the magnetization as $\kappa
\rightarrow \kappa_{c_{1}}$.  The transition at $\kappa =
\kappa_{c_{1}}$ thus appears to be (very) weakly first-order but we cannot
exclude it being second-order, since we cannot rule out the possibility of a continuous but very steep drop to zero of the on-site magnetization as $\kappa \rightarrow \kappa_{c_{1}}$ from the spiral side of the transition, for the same reasons as enunciated above in connection with our discussion of Fig.\ \ref{EvsAngle} and \ref{angleVSj2}.  We find no evidence at all for any
intermediate phase between the quasiclassical N\'{e}el and spiral
phases.  These results may be compared with those for the same model
of Weihong et al.\cite{We:1999} who used a linked-cluster series
expansion technique.  They found that while a nonzero value of the
N\'{e}el staggered magnetization exists for $0 \leq \kappa \lesssim
0.7$, the region $0.7 \lesssim \kappa \lesssim 0.9$ has zero on-site
magnetization, and for $\kappa \gtrsim 0.9$ they found evidence of
spiral order.  Nevertheless, their results came with relatively large
errors, especially for the spiral phase, and we believe that our own
results are probably intrinsically more accurate than theirs.

As a further indication of the accuracy of our results we show in
Table~\ref{EandM_spiral}
\begin{table}[t]
  \caption{Ground-state energy per spin and magnetic order parameter (i.e., the on-site magnetization) for the spin-1/2 HAF on the square and triangular lattices.  We show CCM results obtained for the $J_{1}$--$J_{2}'$ model with $J_{1}>0$, using the spiral model state in various LSUB$n$ approximations defined on the triangular lattice geometry, for the two cases $\kappa \equiv J_{2}'/J_{1}=0$ (square lattice HAF, $\phi=0$) and $\kappa=1$ (triangular lattice HAF, $\phi=\frac{\pi}{3}$).  We compare our extrapolated ($n \rightarrow \infty$) results using Eqs.\ (\ref{Extrapo_E}) and (\ref{Extrapo_M}) and various sets of LSUB$n$ data with other calculations.}
\label{EandM_spiral}
\begin{tabular}{cccccc} \hline\hline
\multirow{2}{*}{Method} &  {$E/N$} & {$M$}  & & {$E/N$} & {$M$}  \\ \cline{2-3} \cline{5-6}
&   \multicolumn{2}{c}{square ($\kappa=0$)}  & & \multicolumn{2}{c}{triangular ($\kappa=1$}) \\  \hline
LSUB$2$ & -0.64833 & 0.4207 & & -0.50290 & 0.4289  \\ 
LSUB$3$ & -0.64931 & 0.4182 & & -0.51911 & 0.4023  \\      
LSUB$4$ & -0.66356 & 0.3827 & & -0.53427 & 0.3637  \\    
LSUB$5$ & -0.66345 & 0.3827 & & -0.53869 & 0.3479  \\      
LSUB$6$ & -0.66695 & 0.3638 & & -0.54290 & 0.3280 \\             
LSUB$7$ & -0.66696 & 0.3635 & & -0.54502 & 0.3152  \\             
LSUB$8$ & -0.66816 & 0.3524 & & -0.54679 & 0.3018 \\ \hline\hline   
\multicolumn{6}{c}{Extrapolations} \\ \hline
LSUB$\infty$ $^{a}$ & -0.66978 & 0.3148 & & -0.55113 & 0.2219 \\ 
LSUB$\infty$ $^{b}$ & -0.66974 & 0.3099 & & -0.55244 & 0.1893  \\     
LSUB$\infty$ $^{c}$ & -0.67045 & 0.3048 & & -0.55205 & 0.2085  \\ \hline
QMC $^{d,e}$  & -0.669437(5) &  0.3070(3)  & &  -0.5458(1) & 0.205(10) \\ 
SE $^{f,g}$ & -0.6693(1) & 0.307(1) & & -0.5502(4) & 0.19(2) \\ \hline\hline
\end{tabular} 
\begin{flushleft}
$^{a}$ Based on $n=\{2,4,6,8\}$  \\ \protect $^{b}$ Based on $n=\{4,6,8\}$ \\ \protect $^{c}$ Based on $n=\{3,5,7\}$ \\ \protect $^{d}$ QMC (Quantum Monte Carlo) for square lattice\cite{Sa:1997} \\ \protect $^{e}$ QMC for triangular lattice\cite{Ca:1999} \\ \protect $^{f}$ SE (Series Expansion) for square lattice\cite{Zh:1991} \\ \protect $^{g}$ SE for triangular lattice\cite{Zh:2006}
\end{flushleft} 
\end{table}
data for the two cases of the spin-1/2 HAF on the square lattice
($\kappa=0$) and on the triangular lattice ($\kappa=1$).  For both
cases we present our CCM results in various LSUB$n$ approximations
(with $2 \leq n \leq 8$) based on the triangular lattice geometry
using the spiral model state, with $\phi=0$ for the square lattice and
$\phi = \frac{\pi}{3}$ for the triangular lattice.  Results are given
for the gs energy per spin $E/N$, and the magnetic order parameter
$M$.  We also display our extrapolated ($n \rightarrow \infty$)
results using the schemes of Eqs.\ (\ref{Extrapo_E}) and
(\ref{Extrapo_M}) with the three data sets $n=\{2,4,6,8\}$,
$n=\{4,6,8\}$ and $n=\{3,5,7\}$.  The results are seen to be very
robust and consistent.  For comparison we also show the results
obtained for the two lattices using quantum Monte Carlo (QMC)
methods\cite{Sa:1997,Ca:1999} and linked-cluster series
expansions.\cite{Zh:2006,Zh:1991} For the square lattice there is no
dynamic frustration and the Marshall-Peierls sign rule\cite{Ma:1955}
applies, so that the QMC ``minus-sign problem'' may be circumvented.
In this case the QMC results\cite{Sa:1997} are extremely accurate, and
indeed represent the best available for the spin-1/2 square-lattice
HAF.  Our own extrapolated results are in complete agreement with
these QMC benchmark results, as found previously (see, e.g.,
Ref.~[\onlinecite{Fa:2008}] and references cited therein), even though
the LSUB$n$ configurations are defined here on the triangular lattice
geometry.  Thus, we note that whereas the individual LSUB$n$ results
for the spin-1/2 square-lattice HAF do not coincide with previous
results for this model (see, e.g., Ref.~[\onlinecite{Fa:2008}])
because previous results have been based on defining the fundamental
LSUB$n$ configurations on a square-lattice geometry rather than on the
triangular-lattice geometry used here, the corresponding LSUB$\infty$
extrapolations in the two geometries are in complete agreement with
each other.

By contrast, the nodal structure of the gs wave function is not
exactly known for the spin-1/2 triangular-lattice HAF, and the QMC
minus-sign problem cannot now be avoided for such frustrated spin
systems.  The QMC results shown\cite{Ca:1999} for the triangular
lattice in Table~\ref{EandM_spiral} were performed using a Green's
function Monte Carlo method with a fixed-node approximation that was
then relaxed in a controlled but approximate way using a stochastic
reconfiguration technique.  For the triangular lattice case we also
show results in Table~\ref{EandM_spiral} from a large-scale
calculation using a linked-cluster series expansion.\cite{Zh:2006} For
such frustrated systems this method, along with our CCM, is probably
among the most accurate available.  We see that in this case our
results for the gs energy are in good agreement with the series
expansion results, whereas the QMC estimate for the energy is almost
certainly too high, and its quoted error hence erroneous.  Our best
estimate for the sublattice magnetization in this case, $M=0.20 \pm
0.02$, is in complete agreement with the best available by all other
methods.

The good agreement, both with respect to internal consistency checks
using different extrapolations and with respect to other methods, for
the gs properties of both the above models, gives us considerable
confidence in our results for the spin-1/2 $J_{1}$--$J_{2}'$ model for
all values of $\kappa \equiv J_{2}'/J_{1}$.

We also comment on the $\kappa \rightarrow \infty$ (decoupled 
spin-1/2 1D HAF chains) limits of Figs.\ \ref{E} and \ref{M}.  Firstly,
Fig.\ \ref{E} shows that at large $J_{2}'$ the extrapolated energy per
spin approaches the value $E/N=-0.4431J_{2}'$ which is the same as the
exact result\cite{Hu:1938} from the Bethe ansatz
solution.\cite{Be:1931} By contrast, the
extrapolated magnetic order parameter at large $J_{2}'$ seems to
approach a constant value $M \approx 0.10$, by contrast with the exact
value of zero\cite{Be:1931,Hu:1938} in this limit.  We note that the
1D anisotropic $XXZ$ chain with anisotropy parameter $\Delta$ has an
essential singularity for $M \rightarrow 0$ at the isotropic point
$\Delta \rightarrow 0$ and this is extremely difficult to mimic in any
truncated numerical calculation.  We note, however, that in the regime
$1 \lesssim \kappa \lesssim 2$ the order parameter $M$ decreases
almost linearly, and if this linear decrease were to be extended $M$
would become zero at a value $\kappa \approx 3.5$.

\begin{figure}[t]
\epsfig{file=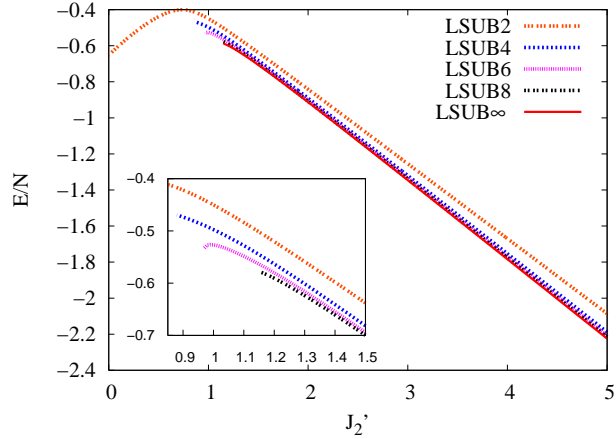,width=6cm,angle=270}
\caption{(Color online) Ground-state energy per spin versus $J_{2}'$
  for the stripe-ordered phase of the spin-1/2 $J_{1}$--$J_{2}'$ Hamiltonian of Eq.\ (\ref{H})
  with $J_{1}=1$.  The CCM results using the
  stripe model state are shown for various LSUB$n$ approximations
  ($n=\{2,4,6,8\}$).  We also show the $n \rightarrow \infty$ extrapolated
  result from using Eq.\ (\ref{Extrapo_E}).}
\label{E_stripe}
\end{figure}

\begin{figure}[b]
\epsfig{file=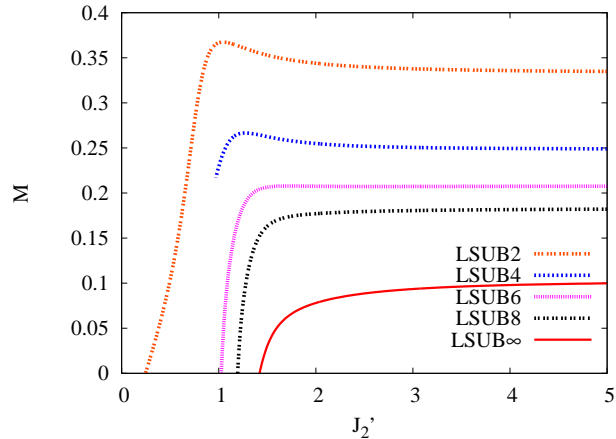,width=6cm,angle=270}
\caption{(Color online) Ground-state magnetic order parameter (i.e.,
  the on-site magnetization) versus $J_{2}'$ for the stripe-ordered
  phase of the spin-1/2 $J_{1}$--$J_{2}'$ Hamiltonian of Eq.\
  (\ref{H}) with $J_{1}=1$.  The CCM results using the stripe model
  state are shown for various LSUB$n$ approximations
  ($n=\{2,4,6,8\}$).  We also show the $n \rightarrow \infty$
  extrapolated result from using Eq.\ (\ref{Extrapo_M}).}
\label{M_stripe}
\end{figure}

We turn finally to our CCM results based on the stripe state as CCM gs
model state $|\Phi \rangle$.  The LSUB$n$ configurations are again
defined with respect to the triangular lattice geometry, exactly as
before.  Results for the gs energy and magnetic order parameter are
shown in Figs.\ \ref{E_stripe} and \ref{M_stripe} respectively for the
collinear stripe phase.
%%%%%%%%%%%%%%%%%%%%
%%%%%%%%%%%%%%%%%%%%%%
They are the precise analogs of Figs.\ \ref{E} and \ref{M} for the
(N\'{e}el and) spiral phases.  We see from Fig.\ \ref{E_stripe} that
some of the LSUB$n$ solutions based on the stripe state show a clear
termination point $\kappa_{t}$ of the sort discussed previously, such
that for $\kappa < \kappa_{t}$ no real solution for the stripe phase
exists.  In particular the LSUB$6$ and LSUB$8$ solutions terminate at
the values shown in Table~\ref{spiralVSstripe_all}.
\begin{table}[t]
%\begin{center}
\vskip0.5cm
\caption{The parameters $\kappa_{e}$ (the crossing point of the energy
  curves for the stripe and spiral phases) and $\kappa_{t}$ (the
  termination point of the stripe state solution) in various LSUB$n$
  approximations defined on the triangular lattice geometry, for the
  spin-1/2 $J_{1}$--$J_{2}'$ model, with $\kappa \equiv J_{2}'/J_{1}$,
  $J_{1}>0$.  The ``LSUB$\infty$'' extrapolations are explained in the
  text.}
\label{spiralVSstripe_all}
\begin{tabular}{ccc}    \hline\hline
\multirow{2}{*}{LSUB$n$}  & \multicolumn{2}{c}{$J_{2}'$} \\ \cline{2-3}
 & $\kappa_{e}$ & $\kappa_{t}$  \\ \hline
LSUB2 & $\infty$ &  - \\ 
LSUB4 & 4.555 & (0.880) \\ 
LSUB6 & 3.593 & 0.970 \\ 
LSUB8 & 3.125 & 1.150 \\    
``LSUB$\infty$'' & $1.69 \pm 0.03$ & 1.69  \\ \hline\hline  
\end{tabular}
\end{table}
As is often the case the LSUB$2$ solution does not terminate, while
the LSUB$4$ solution shows a marked change in character around the
value $\kappa \approx 0.880$ that is not exactly a termination point
(but, probably, rather reflects a crossing with another unphysical
solution).  In any event, the LSUB$4$ data are not shown below this
value in Figs.\ \ref{E_stripe} and \ref{M_stripe}.

\begin{figure}[b]
\epsfig{file=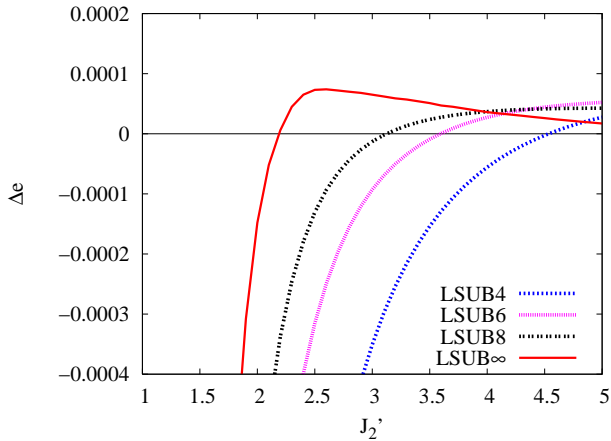,width=6cm,angle=270}
\caption{(Color online) Difference between the ground-state energies per spin ($e \equiv E/N$) of the spiral and stripe phases ($\Delta e \equiv e^{{\rm spiral}}- e^{{\rm stripe}}$) versus $J_{2}'$ for the spin-1/2
  $J_{1}$--$J_{2}'$ Hamiltonian of Eq.\ (\ref{H}) with $J_{1}=1$.  The
  CCM results for the energy
  difference using the stripe and spiral model states for various LSUB$n$ approximations (n$=\{4,6,8\}$) are
  shown.  We also show the $n \rightarrow \infty$ extrapolated result
  from using Eq.\ (\ref{Extrapo_E}) for the two phases separately.}
\label{E_diff}
\end{figure}

The large $\kappa$ limit of the energy per spin results of Fig.\
\ref{E_stripe} again agrees with the exact 1D chain result of
$E/N=-0.4431J_{2}'$, just as in Fig.\ \ref{E} for the spiral phase.
However, the most important observation is that for all LSUB$n$
approximations with $n>2$ the curves for the energy per spin of the
stripe phase cross with the corresponding curves (i.e., for the same
value of $n$) for the energy per spin of the spiral phase at a value
that we denote as $\kappa_{e}$.  Thus, for $\kappa < \kappa_{e}$ the
spiral phase is predicted to be the stable phase (i.e., lies lowest in
energy), whereas for $\kappa > \kappa_{e}$ the stripe phase is
predicted to be the stable ground state.  We thus have a clear first
indication of another (first-order) quantum phase transition in the
spin-1/2 $J_{1}$--$J_{2}'$ model at a value $\kappa = \kappa_{c_{2}}$.
Figure~\ref{E_diff}
shows the energy difference between the stripe and spiral states for
various LSUB$n$ calculations, and some indicative values near the
crossing values $\kappa_{e}$ are also shown in
Table~\ref{spiralVSstripe_all}.

As we remarked in Sec.~\ref{model_section}, the stripe phase is never
the stable classical ground state since it always lies higher in
energy than the spiral phase.  Indeed, it is easy to show that the
classical gs energy per spin, written as $e\equiv
E(Ns^{2}J_{2}')^{-1}$, has the value $e^{{\rm stripe}}_{{\rm cl}}=-1$
(independent of $J_{1}$) for the stripe phase, and $e^{{\rm
    spiral}}_{{\rm cl}}=-1-\frac{1}{2}\kappa^{-2}$ for the spiral
phase in the regime $\kappa \geq \frac{1}{2}$ where the spiral phase
exists classically.  Thus, the difference in energy between the two
phases classically is $e^{{\rm spiral}}_{{\rm cl}} - e^{{\rm
    stripe}}_{{\rm cl}} \equiv \Delta e_{{\rm cl}} =
-\frac{1}{2}\kappa^{-2} < 0$.  By contrast, what we have shown at the
quantum level is that quantum fluctuations can change the sign of
$\Delta e$ at some critical value $\kappa = \kappa_{c_{2}}$ such that
the collinear stripe phase becomes stabilized for all $\kappa >
\kappa_{c_{2}}$.  However, the energy differences $\Delta e$ in this
regime are found to be extremely small, as may be seen from
Fig.\ \ref{E_diff},
\begin{table}[t]
\begin{center}
\vskip0.5cm
\caption{Comparision of the gs energy per spin of the stripe and spiral phases
  of the spin-$1/2$ $J_{1}$--$J_{2}'$ model in various LSUB$n$
  approximations ($n=\{2,4,6,8\}$) at some illustrative values of
  $J_{2}'$ (with $J_{1}=1$).}
\label{spiralVSstripe_E}
\begin{tabular}{cccc}    \hline\hline
\multirow{2}{*}{LSUB$n$} & \multirow{2}{*}{$J_{2}'$} & \multicolumn{2}{c}{$E/N$} \\ \cline{3-4}  
 & & stripe & spiral \\ \hline 
2 & 10 & -4.167865 & -4.168525 \\ 
4 & 4.4 & -1.927326 & -1.927337 \\ 
4 & 4.6 & -2.014224 & -2.014219 \\ 
4 & 4.8 & -2.101154 & -2.101136 \\  
6 & 3.4 & -1.508386 & -1.508406 \\ 
6 & 3.6 & -1.595632 & -1.595632 \\ 
6 & 3.8 & -1.682968 & -1.682952 \\ 
8 & 2.9 & -1.295557 & -1.295584 \\
8 & 3.1 & -1.382665 & -1.382667 \\  
8 & 3.3 & -1.469937 & -1.469924 \\ \hline\hline
\end{tabular}
\end{center}
\end{table}
and hence the stripe phase is predicted to be very fragile against
small perturbations or thermal fluctuations, for example.
Nevertheless, we should stress that the LSUB$n$ energy differences
displayed in Fig.\ \ref{E_diff} are well within (by several orders of
magnitude) the margins of error in our individual calculations.

Clearly, the LSUB$n$ energy crossing points $\kappa_{e}(n)$ at which
$\Delta e_{{\rm LSUB}n} = 0$ provide a measure of $\kappa_{c_{2}}$.
In the past we have found that a simple linear extrapolation, $\kappa
= c_{0}+c_{1}n^{-1}$, yields a good fit to such critical points, and
this seems to be the case here too.  The corresponding
``LSUB$\infty$'' estimate from the $\kappa_{e}$ LSUB$n$ data of
Table~\ref{spiralVSstripe_all} with $n=\{4,6,8\}$ gives an estimate
$\kappa_{c_{2}} \approx 1.69 \pm 0.03$, where the error is the
standard deviation in the fit.  A similar linear extrapolation on the
LSUB$n$ stripe-phase termination points $\kappa_{t}$ with $n=\{6,8\}$
gives a second estimate $\kappa_{c_{2}} \approx 1.69$, in remarkably
good agreement with the first estimate.  We note too that we can, of
course, also obtain another estimate of $\kappa_{e}$ from the crossing
point of the two extrapolated (LSUB$\infty$) gs energy curves for the
spiral and stripe-ordered phases (i.e., using Eq.\ (\ref{Extrapo_E})
in each case), shown in Figs.\ \ref{E} and \ref{E_stripe}
respectively, and it is this LSUB$\infty$ result that is displayed in
Fig.\ \ref{E_diff}.  Since the two curves cross at such a very shallow
angle, and since both curves are extrapolated completely independently
of one another, we expect that this estimate for $\kappa_{e}$ is
perhaps intrinsically less accurate than the one discussed above.
Nevertheless, it is very gratifying that the value so obtained for the
crossing point of the two extrapolated (LSUB$\infty$) gs energy
curves, namely $\kappa_{e} \approx 2.19$, is rather close to the
previous value, considering that the two curves are almost parallel to
one another in the crossing region.  The difference in the two
estimates for $\kappa_{e}$ of about 1.7 and 2.2 is itself an
indication of the error in these estimates in this incredibly
difficult regime where the two phases lie so close in energy to one
another.  Nevertheless, we reiterate that the CCM retains sufficient
accuracy for both the spiral and stripe phases individually (both in
the raw LSUB$n$ data and in the extrapolated LSUB$\infty$ results for
each phase) to ensure that, even though the corresponding estimates
for the energy difference $\Delta e$ are small in the crossing regime
and for values of $\kappa > \kappa_{c}$, they are still sufficiently
large to be well within our limits of accuracy.

Finally, from Fig.\ \ref{M_stripe} we see the corresponding results for
the magnetic order parameter of the stripe phase.  The large-$\kappa$
limit of uncoupled 1D spin-1/2 chains is identical to that for the
spiral phase, and hence suffers from the same (known) problem of not
giving the correct $M=0$ result in this limit.  However at smaller
values of $\kappa$ the order parameter decreases and becomes zero at
some critical value $\kappa_{m}(n)$ that depends on the LSUB$n$
approximation.  The extrapolated ($n \rightarrow \infty$) result
obtained in the usual way from Eq.\ (\ref{Extrapo_M}) is also shown in
Fig.\ \ref{M_stripe} and is seen to become zero at a value $\kappa_{m}
\approx 1.43$.  Since the phase transition at $\kappa =
\kappa_{c_{2}}$ is clearly of first-order type from the energy data,
the magnetization data provide us only with an inequality
$\kappa_{c_{2}} \geq \kappa_{m}$.  

In summary, although it is difficult to put firm error bars on our
results for our predicted second critical point, our best current
estimate, based on all the above results, is $\kappa_{c_{2}} = 1.8 \pm
0.4$.

\section{Discussion and Conclusions}
\label{discussion}
In this paper we have used the CCM to study the influence of quantum
fluctuations on the zero-temperature gs phase diagram of a spin-half
Heisenberg antiferromagnet, the $J_{1}$--$J_{2}'$ model, defined on an
anisotropic 2D lattice.  We have studied the case where the NN $J_{1}$
bonds are antiferromagnetic ($J_{1} > 0$) and the competing $J_{2}'
\equiv \kappa J_{1}$ bonds have a strength $\kappa$ that varies from
$\kappa = 0$ (corresponding to the spin-half HAF on the square
lattice) to $\kappa \rightarrow \infty$ (corresponding to a set of
decoupled spin-half 1D HAF chains), with the spin-half HAF on the
triangular lattice as the special case $\kappa = 1$ in between the two
extremes.

Whereas at the classical level the model has only two stable gs
phases, one with N\'{e}el order for $\kappa < \kappa_{{\rm cl}} = 0.5$
and another with spiral order for $\kappa > \kappa_{{\rm cl}}$, the
quantum fluctuations for the spin-half model can stabilize a third
non-classical phase with collinear stripe ordering at sufficiently
high values of $\kappa$.  Thus, we find two quantum phase transitions,
both seemingly first-order.  The first at $\kappa_{{\rm cl}} = 0.80
\pm 0.01$ separates a phase with classical N\'{e}el ordering for
$\kappa < \kappa_{c_{1}}$ from a phase with helical ordering
for $\kappa > \kappa_{c_{1}}$.  This latter phase includes the case
$\kappa = 1$ of the spin-half triangular-lattice HAF with the standard
120$^{\circ}$ three-sublattice quasiclassical ordering.  By contrast
with the classical second-order transition at $\kappa_{{\rm cl}}=0.5$,
the quantum phase transition at $\kappa_{c_{1}} \approx 0.80$ appears
to be weakly first-order in nature, although we cannot exclude
it from being second-order.  The N\'{e}el order thus survives into a
region $0.5 < \kappa < \kappa_{c_{1}}$ where it is classically
unstable.  This is an example, among many others, of the widely
observed phenomenon that quantum fluctuations tend to favor collinear
ordering.

A second quantum phase transition is predicted by our calculations to
occur at $\kappa_{c_{2}}=1.8 \pm 0.4$.  It separates the 
helical phase for $\kappa < \kappa_{c_{2}}$ from a phase with
collinear stripe ordering at $\kappa > \kappa_{c_{2}}$.  However in
this latter region, $\kappa > \kappa_{c_{2}}$, the stripe and spiral
phases are extremely close in energy, and hence the stripe phase may
be very fragile against external perturbations or thermal fluctuations,
for example.  The existence of the collinear stripe phase seems to
rely again on the fact that quantum fluctuations favor collinear
ordering.  An alternative, but essentially equivalent, explanation
starts by looking at the large-$\kappa$ limit of uncoupled 1D HAF
chains, for which all states with spins on different chains orienting
randomly with respect to each other are degenerate in energy in the
$\kappa \rightarrow \infty$ limit.  Then, as the interchain coupling
$J_{1}=J_{2}'/\kappa$ is slowly increased from zero, what seems to
occur is that the relative orientations are locked into collinearity
by the familiar phenomenon of order by disorder.\cite{Vi:1977} As a
somewhat technical aside at this point we note, however, that while
the present $J_{1}$--$J_{2}'$ model attains (collinear) order from an
otherwise disordered set of spin chains, the order by disorder
phenomenon seemingly responsible here differs in an important respect
from its archetypal realization in the $J_{1}$--$J_{2}$ model.  In
this latter case each of the inter-penetrating sublattices is
characterized by a classical magnetization vector, and the classical
disorder emanates from the independence of the energy on the mutual
orientation of the magnetization vectors for the two sublattices.  By
contrast, for the present $J_{1}$--$J_{2}'$ model, the corresponding
phenomenon is more subtle since the spin chains that now order
(collinearly) are themselves quantum-critical.  Hence, what is
established now is not only that the relative orientations of the
magnetizations on different chains become locked into collinear order,
but also the very fact that such a classical notion is actually
appropriate.  

We note that our own calculations are weakest precisely in the
$\kappa \rightarrow \infty$ limit where, although our result for the
gs energy of the chains is excellent, the subtle quantum-critical
ordering of the Bethe ansatz solution that results in the value $M=0$
for the staggered magnetization,\cite{Hu:1938} is not exactly
reproduced.  Nevertheless, with the single exception of our inability
to reproduce exactly this very singular and non-analytic result for
the on-site magnetization,\cite{Hu:1938} the CCM calculations are
remarkably robust and accurate over the rest of the parameter space.
Thus, our results for the gs energy and on-site magnetization have
provided a set of independent checks that lead us to believe that we
have a self-consistent and coherent description of this interesting
model.

Before concluding we compare our results with those from previous
calculations on the same model.  The simplest such studies have
utilized lowest-order (or linear) spin-wave theory
(LSWT)\cite{Tr:1999,Me:1999} and Schwinger boson mean-field theory
(SBMFT).\cite{Ga:1993} The independent LSWT calculations of
Trumper\cite{Tr:1999} and Merino {\it et al.}\cite{Me:1999} both found
a continuous second-order phase transition from a N\'{e}el-ordered to
a spiral-ordered phase at precisely the classical value
$\kappa_{c_{1}}=0.5$, at which the magnetization order parameter
approaches zero continuously from both sides.  Although LSWT is known
to give a reasonable description of the spin-1/2 Heisenberg
antiferromagnet on both the square lattice ($\kappa=0$) and the
triangular lattice ($\kappa=1$), it is clearly unable to model the
intermediate regime accurately.  This is particularly true around the
point of maximum classical frustration at $\kappa=1/2$, for which one
fully expects, as mentioned previously, that N\'{e}el order is
preserved to higher values of $\kappa$ than pertain classically, as
found both by earlier more accurate studies, such as those using
series expansion (SE) techniques,\cite{We:1999} and by us in the
present paper.

Similar shortcomings of spin-wave theory (SWT) have been noted by
Igarashi\cite{Ig:1993} in the context of the related spin-1/2
$J_{1}$--$J_{2}$ model on the square lattice, discussed briefly in
Sec.\ \ref{Introd}.  He showed that whereas its lowest-order version
(LSWT) works well when $J_{2}=0$, it consistently overestimates the
quantum fluctuations as the frustration $J_{2}/J_{1}$ increases.  In
particular he showed by going to higher orders in SWT in powers of
$1/s$ where $s$ is the spin quantum number and LSWT is the leading
order, that the expansion converges reasonably well for $J_{2}/J_{1}
\lesssim 0.35$, but for larger values of $J_{2}/J_{1}$, including the
point $J_{2}/J_{1}=0.5$ of maximum classical frustration, the series
loses stability.  He also showed that the higher-order corrections to
LSWT for $J_{2}/J_{1} \lesssim 0.4$ make the N\'{e}el-ordered phase
more stable than predicted by LSWT.  He concluded that any predictions
from SWT for the spin-1/2 $J_{1}$--$J_{2}$ model on the square lattice
are likely to be unreliable for values $J_{2}/J_{1} \gtrsim 0.4$.  It
is likely that a similar analysis of the SWT results for the present spin-1/2
$J_{1}$--$J_{2}'$ model on the square lattice would reveal similar shortcomings of LSWT as the frustration parameter $\kappa \equiv J_{2}'/J_{1}$ is increased.

By contrast with the above LSWT results for the spin-1/2
$J_{1}$--$J_{2}'$ model on the square lattice, a SBMFT
analysis\cite{Ga:1993} shows a continuous transition from a collinear
N\'{e}el phase to a spiral phase at a value $\kappa_{c_{1}} \approx
0.6$, but with a non-varnishing magnetization, $M \approx 0.175$, at
the critical point.  The discrepancy with our own results (viz.,
$\kappa_{c_{1}} \approx 0.80$ with either a vanishing or very small
magnetization, $M \approx 0.025 \pm 0.025$, at the critical point) is
almost certainly again due to the lowest-order nature of the
mean-field approach, and particularly the complete neglect at the
SBMFT level of Gaussian fluctuations.

Turning to the second phase transition we note that LSWT
predicts that the magnetization in the spiral phase of the present
spin-1/2 $J_{1}$--$J_{2}'$ model vanishes at a value $\kappa \approx
3.70$.  The authors of these results\cite{Tr:1999,Me:1999} took that
to indicate the possible existence of a disordered phase for $\kappa
\gtrsim 3.70$.  Even ignoring the probably unreliable nature of LSWT
results for such high values of $\kappa$, as noted above, the safer
conclusion based on them is that the spiral phase simply becomes
unstable for $\kappa \gtrsim 3.70$.  Indeed, SWT results are always
based on a particular choice of phase, usually based on a classical
($s \rightarrow \infty$) analog, just as are our own CCM calculations.
In both cases the vanishing of an order parameter only signals a phase
transition to another state, but in the absence of another calculation
based on another (classical) hypothesized phase, no conclusion about
the adjacent phase can be drawn.  It is precisely such an additional
calculation based on a stripe-ordered phase that has been done in the
present work, which has shown the onset of this phase (at a
lower-energy than the spiral phase) for $\kappa > \kappa_{c_{2}}
\approx 1.8 \pm 0.4$.  No such calcultions were attempted in either the
LSWT\cite{Tr:1999,Me:1999} or the SBMFT\cite{Ga:1993} cases, and hence
no direct comparison can be drawn with own results for this second
phase transition, except to say that the LSWT result with its expected
uncertainties discussed above is not incompatible with our own
conclusions.

As has been noted elsewhere,\cite{Bi:2008_JPCM} high-order CCM results
of the sort presented here are believed to be among the best available
for such highly frustrated spin-lattice models.  Many previous
applications of the CCM to unfrustrated spin models have given
excellent quantitative agreement with other numerical methods
(including exact diagonalization (ED) of small lattices, quantum Monte
Carlo (QMC), and series expansion techniques).  A typical example is
the spin-half HAF on the square lattice, which is the $\kappa=0$ limit
of the present model (and see Table~\ref{EandM_spiral}).  It is
interesting to compare for this $\kappa=0$ case, where comparison can
be made with QMC results, the present CCM extrapolations of the
LSUB$n$ data for the infinite lattice to the $n \rightarrow \infty$
limit and the corresponding QMC or ED extrapolations for the results
obtained for finite lattices containing $N$ spins that have to be
carried out to give the $N \rightarrow \infty$ limit.  Thus, for the
spin-1/2 HAF on the square lattice the ``distance'' between the CCM
results for the ground-state energy per spin\cite{Fa:2008} at the
LSUB8 (LSUB10) level and the extrapolated LSUB$\infty$ value is
approximately the same as the distance of the corresponding QMC
result\cite{Ru:1992} for a lattice of size $N=12 \times 12$ ($N=16
\times 16$) from its $N \rightarrow \infty$ limit.  The corresponding
comparison for the magnetic order parameter $M$ is even more striking.
Thus even the CCM LSUB6 result for $M$ is closer to the LSUB$\infty$
limit than any of the QMC results for $M$ for lattices of $N$ spins
are to their $N \rightarrow \infty$ limit for all lattices up to size
$N=16 \times 16$, the largest for which calculations were
undertaken.\cite{Ru:1992} Such comparisons show, for example, that
even though the ``distance'' between our LSUB$n$ data points for $M$
and the extrapolated ($n \rightarrow \infty$) LSUB$\infty$ result
shown in Fig.\ \ref{M} may, at first sight, appear to be large, they
are completely comparable to or smaller than those in alternative
methods (where they can be applied).  Furthermore, where such
alternative methods can be applied, as for the spin-1/2 HAF on the
square lattice, the CCM results are in complete agreement with them.

By contrast, for frustrated spin-lattice models in two dimensions both
the QMC and ED techniques face formidable difficulties.  These arise
in the former case due to the ``minus-sign problem'' present for
frustrated systems when the nodal structure of the gs wave function is
unknown, and in the latter case due to the practical restriction to
relatively small lattices imposed by computational limits.  The latter
problem is exacerbated for incommensurate phases, and is compounded
due to the large (and essentially uncontrolled) variation of the
results with respect to the different possible shapes of clusters of a
given size.

Thus, for highly frustrated spin-lattice models like the present
$J_{1}$--$J_{2}'$ model, the best alternative numerical method to the
CCM is the linked-cluster series expansion (SE)
technique.\cite{We:1999,Pa:2008,He:1990,Ge:1990,Ge:1996} The SE
technique has also been applied to the present
model.\cite{We:1999,Pa:2008} The earlier study\cite{We:1999} mainly
dealt with the N\'{e}el and spiral phases.  Unlike in that work we
find no evidence at all for an intermediate (dimerized) phase between
the N\'{e}el and spiral phases in the parameter regime $0.7 \lesssim
\kappa \lesssim 0.9$.  The very recent SE study\cite{Pa:2008} was
motivated by the prediction of Starykh and Balents\cite{St:2007} for
the existence of a stable collinear stripe-ordered gs phase for values
of $\kappa$ above some critical value that they did not calculate.
The SE study showed that although the collinear stripe phase was
stabilized for large values of $\kappa$ relative to the classical
result, nevertheless in their calculations the non-collinear helical
phase was still always lower in energy.  Hence, they could not confirm the
existence of the stripe-ordered phase.  They concluded by suggesting
that further unbiased ways of studying the competition between the
spiral and stripe phases would be useful.  We believe that the present
CCM calculations provide exactly such unbiased results, which now do
indeed appear to confirm the prediction of Starykh and Balents.

We end by remarking that it would also be of interest to repeat the
present study for the case of the spin-one $J_{1}$--$J_{2}'$ model.
The calculations for this case are more demanding due to an increase
at a given LSUB$n$ level of approximation in the number of fundamental
configurations retained in the CCM correlation operators.
Nevertheless, we hope to be able to report results for this system in
the future.

\section*{ACKNOWLEDGMENTS}
We thank the University of Minnesota Supercomputing Institute for
Digital Simulation and Advanced Computation for the grant of
supercomputing facilities, on which we relied heavily for the
numerical calculations reported here.

\end{document}